\documentclass[preprint, 12pt]{aastex}

\slugcomment{Accepted to AJ}

\begin{document}


\title{Forbidden Line Emission in the Eccentric Spectroscopic Binaries DQ Tauri and UZ Tauri E Monitored over an Orbital Period}

\author{M. Huerta and P. Hartigan}
\affil{Physics and Astronomy Department, Rice University, Houston, TX 77005; \email {marcosh@rice.edu, hartigan@sparky.rice.edu}}

\author{R. J. White\altaffilmark{1}}
\affil{Department of Astronomy, University of Texas at Austin, Austin, TX, 78712}

\altaffiltext{1}{Current address: Department of Astronomy, 105-24 Caltech, 1201 East California Blvd, Pasadena, CA 91125}

\begin{abstract}

We present echelle spectroscopy of the close pre-main-sequence binary star systems DQ Tau and UZ Tau-E.  Over a 16 day time interval we acquired 14 nights of spectra for DQ Tau and 12 nights of spectra for UZ Tau-E.  This represents the entire phase of DQ Tau, and 63 percent of the phase of UZ Tau-E.  As expected, photospheric lines such as Li I $\lambda$6707 clearly split into two components as the primary and secondary orbit one another, as did the permitted line He I $\lambda$5876.  Unlike the photospheric features, the forbidden lines of [O~I] $\lambda$6300 and [O~I] $\lambda$5577, retain the same shape throughout the orbit.  Therefore these lines must originate outside of the immediate vicinity of the two stars and any circumstellar disks that participate in the orbital motion of the stars.

\end{abstract}

\newpage

\section{Introduction}

Forbidden emission lines are important diagnostics of the mass loss from young stellar objects (Cabrit et al. 1990; Hartigan, Edwards, \& Ghandour 1995).  In young stars, the forbidden lines are strong in low-density mass outflows, and are typically blue-shifted as the red-shifted component is obscured by a circumstellar disk.  Spectroscopic surveys \citep[e.g.][]{hir97, sol99, pyo03} have shown that forbidden lines in T Tauri stars sometimes have double peaked velocity profiles, with both a high-velocity component (HVC) on the order of hundreds of km/s and a low-velocity component (LVC) closer to the rest velocity of the star.  The surveys also demonstrate that forbidden lines have components that extend greatly from the star and  also that only emit from very near the star.  In the stars with clear double peaked forbidden line profiles, the HVC extends outward on arcsecond scales, while the LVC does not, as pointed out by \citet{hir97}.

The difficulty in determining the origins of the near-star forbidden line emission exists because it is offset not more than a few tenths of an arcsecond from the star \citep{hir97}.  The source region may well lie within a few stellar radii of the source, and would require a spatial resolution requirement of milliarcseconds to observe directly, not possible with present technology.  Further, young stars form in regions where dust, gas, and other sources of light can obscure the star itself from view, so it is often difficult to follow embedded jets all the way to the star \citep{eis00}.

Fortunately, most stars are in binary systems, and we can use the kinematics of binary T Tauri stars to indirectly probe the origin of the forbidden lines that are not extended from the star.  Highly eccentric binary stars yield large Doppler shifts that appear in any emission or absorption features that participate in the orbital motion of the two stars.  If the young binary stars have circumstellar disks, and the origin of the near forbidden lines is linked to the disks, then we can expect to see orbital variation of the forbidden line profiles in tandem with line profile changes in photospheric features of the stars themselves.

DQ Tau and UZ Tau-E are pre-main sequence systems with signatures of active accretion disks, i.e. classical T-Tauri stars.  Both systems are spectroscopic binaries;  DQ Tau has a period of 15.8 days \citep{mat97} and UZ Tau-E a period of 19 days \citep{pra02}.  Each system has a highly eccentric orbit, with spectra containing strong forbidden lines.  It is worth noting that the forbidden line profiles of DQ Tau and UZ Tau-E lack distinct dual-velocity components.  Indeed, DQ Tau does not have much of an HVC at all.  UZ Tau-E did have a somewhat noticeable double peaked profile in \citet{har95}, and also has a microjet of 0.2 arcseconds observed in [O~I] 6300 slitless imaging (Hartigan, Edwards, \& Pierson 2004). The short periods and high eccentricities of these two systems made them ideal candidates for localizing the near-star emission with the method outlined above.

\section {Observations and Data Reduction}

Observations were taken with the echelle spectrograph on the 2.1-m Otto Struve telescope at McDonald Observatory from January 4 - 19, 2002.  Using the RA2 CCD at 1232x400 with a 1" by 8" slit we obtained R = 46,000 spectra from 5460 \AA{} to 6760 \AA{}.  The weather was excellent, and we obtained 14 nights of spectra (Table 1).  However, one of the lost days was near the close approach of the DQ Tau system and therefore at the point were stellar spectral features would be most split.

DQ Tau was typically observed in 20 minute exposures for a total of 2 hours each night, and sometimes more when time and weather permitted.  UZ Tau-E was observed for about one hour each night (Table 1).  The UZ Tau system also contains a visual binary, UZ Tau-W, approximately 3.5" to the west of UZ Tau-E;  the slit was oriented such that UZ Tau-W did not contaminate the spectrum of UZ Tau-E.  In addition, the weak line T-Tauri star V819 Tau was observed for photospheric fitting and spectra were taken of the bright O-type star, S Mon, to correct for telluric absorption which affects the line profiles around [O~I] $\lambda$6300.  To assist in the removal of night sky emission lines, the observations were conducted in January when the annual velocity of the Earth ($\sim$ -20 km/s, with respect to Taurus) generates the largest velocity shift between telluric features and the systemic velocities of UZ Tau E and DQ Tau.  We smoothed all of the spectra presented in this paper with a moving average of $5-7$ pixels, corresponding to about $5-7$  km~s$^{-1}$ to improve the signal-to-noise of the reduced data.

We reduced the raw data using the echelle data reduction routines in IRAF\footnote{IRAF (Image Reduction and Analysis Facility) is distributed by the National Optical Astronomy Observatories, which are operated by the Association of Universities for Research in Astronomy (AURA), Inc., under cooperative agreement with the National Science Foundation}.  The raw data files were bias-subtracted and flat fielded before spectral extraction.  We traced the echelle orders using the \texttt{doecslit} package.  Wavelength calibration was done with a Thorium-Argon lamp and with spectral values taken from the NOAO spectral database.

The spectra of V819 Tau, a weak-line T Tauri star of the same spectral type as DQ Tau, was used to subtract photospheric features from DQ Tau and improve the line profiles.  To create the photospheric features of a spectroscopic binary, two appropriately Doppler shifted spectra of V819 Tau were added to each other.   This technique successfully reproduced the photospheric features of DQ Tau.  Absorption features in T Tauri stars can be masked by excess continuum from material accreting onto the star, a process known as veiling.  We accounted for veiling by altering the strengths in the absorption features of the two templates by adding a constant and re-normalizing to the continuum.  The process of adding continuum to the template stars accounts for veiling in DQ Tau.  The technique produced good photospheric fits in the echelle orders of interest which when subtracted from DQ Tau succeeded in removing lines from the photosphere (see Figure 1).

We subtracted the night sky lines for [O~I] $\lambda$6300 and [O~I] $\lambda$5577 by using two extraction window sizes, one narrow and one wide.  By subtracting, with appropriate scaling, the narrow window spectrum from the wide window spectrum, it was possible to create a profile of only the night sky line which could then be used for subtraction.  We replaced the original order with the sky-subtracted order in only the wavelength range affected by the night sky line, with the sky-subtracted segment scaled to fit the original spectrum.  This technique was very successful in producing night-sky subtracted spectrum for the [O~I] $\lambda$6300 line (Figure 2).  The same technique was used to remove the night sky in the [O~I] $\lambda$5577 line , which had brighter night sky emission and weaker emission from the object, creating a noisier residual line profile than [O~I] $\lambda$6300. 

\section{Results}

\subsection{Orbital Elements}

The observing run yielded fourteen nights of spectra, including the forbidden lines of interest and other prominent photospheric features with which we could trace the orbital motion of the primary and secondary.  Only eleven of the fourteen spectra were used in measuring radial velocities of the secondary and primary for purposes of comparison to the orbital elements.   The radial velocities measured in our data came from measuring shifts in Li I $\lambda$6707,  Ca I $\lambda$6717.7, and Al I $\lambda$6696.02, which you can see in Figure 3.  The vertical error bars represent one standard deviation in the average of radial velocities measured from double-gaussian fits to those three absorption lines.  The radial velocity measurements allowed us to adjust slightly the period found in  \citet{mat97}, which gave P = 15.8043 $\pm$ 0.0024 days.  Our observations were 170 epochs since the time of periastron passage, T = 2449582.54 $\pm$ 0.05 JD.  We found the best fit between our data and the orbit of \citet{mat97} came when we shortened the period to 15.8016 $\pm^{0.002}_{0.006}$ days. 

\subsection{Variations of Line Emission with Orbital Phase}

Signal for the [O~I] $\lambda$6300 line in DQ Tau was very good throughout the run and the line profile remains unchanged as the stellar velocities approach their maximum, as shown in Figure 4.  The permitted line He I $\lambda$5876 both splits and shows intensity variation as previously observed \citep{bas97}.  UZ Tau-E also shows no variation in the [O~I] $\lambda$6300 line (Figure 5).  The weakness of [O~I] $\lambda$5577 makes any changes in its line profile difficult to measure in DQ Tau, and impossible in UZ Tau-E.  To improve our signal for DQ Tau, we co-added five nights of data near periastron, and compared the result with the sum of five nights far from periastron (see Figure 6.)  The [O~I] lines again lack any variation with orbital phase.

Forbidden line profile variation with orbital phase, or lack thereof, was previously unknown for any spectroscopic T Tauri binary.  Since the HVC is typically associated with emission that is greatly extended from the star, the lack of variation in the HVC would be unsurprising.  However, in DQ Tau the HVC is not an obvious feature and the entire profile is mostly at low velocity,  showing no change with orbital phase.  A double peaked profile of [O~I] $\lambda$6300 is better seen in Figure 5 for UZ Tau-E, which has a clearer distinction between the LVC (around 6301 \AA{}) and the HVC (around 6299 \AA{}), and also shows no variation with phase.  The two velocity components of UZ Tau-E are somewhat less distinct than the double peak profile in \citet{har95}.

\section{Discussion}

The [O~I] $\lambda$6300 line in DQ Tau and UZ Tau-E shows no variation with orbital phase.  While the signal strength is not high for [O I] $\lambda 5577$, the multi-day averages in Figure 6 show no significant changes between high-velocity and low velocity observations.  The data indicates that in both systems, the [O~I] $\lambda$6300 emission must not come from any source associated with the individual stars, such as a circumstellar disk or a disk wind in which the emission comes from close to the stars \citep{har95, kwa95}.   

It is possible that an LVC could originate in a disk wind and still not vary with orbital phase.  If a disk wind driven outflow extends beyond ~0.2 AU then forbidden line emission would be averaged out in velocity over multiple orbits.   The distance is estimated from a typical LVC speed of ~25 km/s multiplied by the $\sim$14 day orbital period of DQ Tau.  The LVC in a disk wind can originate out to $\gtrsim 3$ AU from the star \citep{kwa97}, a distance more than sufficient to average out any profile changes that occur due to orbital motion.

We have provided an additional observational constraint on the formation of forbidden lines in classical T-Tauri stars.  We can eliminate any process that would tie the emission spatially to the stars or their circumstellar disks.  The LVC could be produced as part of the outflow, for example, as wakes of bow shocks in a jet that intersects the edges of an evacuated cavity near the star, or in a disk wind at $\gtrsim 0.2$ AU.  The LVC could also arise from a circumbinary disk that does not partake of the orbital motion of the component stars.

\acknowledgments

The authors would like to thank Chris Johns-Krull for his help with the radial velocitiy plot of DQ Tau, and for assistance in identifying photospheric lines in DQ Tau.  We are grateful to Jennifer Hoffman, for her help and comments on early drafts of this paper.  M.H. was partially supported by NSF Cooperative Agreement Number HRD-9817555 as a part of the Rice University Alliance for Graduate Education and the Professoriate (AGEP) Program.

\begin{deluxetable}{lcrcr}

\tablewidth{0 pt}
\tablecaption{Observation Log}
\tablehead{
\colhead{} & \multicolumn{2}{c}{DQ Tau} & \multicolumn{2}{c}{UZ Tau-E}  \\
\colhead{UT Date}           & 
\colhead{JD - 2452200} & \colhead{Phase\tablenotemark{a}} & \colhead{JD - 2452200} & \colhead{Phase}
}
\startdata

2001 Jan 4 &  78.62 &  0.592 & \nodata & \nodata \\
2001 Jan 5 &  79.61 &  0.654 & \nodata & \nodata \\
2001 Jan 7 &  81.60 &  0.780  & 81.74 & 0.938 \\
2001 Jan 8 &  82.55 &  0.840  & 82.76 & 0.991\\
2001 Jan 9 &  83.56 &  0.904 & 83.66 & 0.039 \\
2001 Jan 10 &  84.57 &  0.968 & 84.71 & 0.093\\
2001 Jan 12 &  86.56 &  0.094 & 86.66 & 0.196\\
2001 Jan 13 &  87.55 &  0.157 & 87.65 & 0.248\\
2001 Jan 14 &  88.55 &  0.220 & 88.68 & 0.302\\
2001 Jan 15 &  89.58 &  0.285 & 89.69 & 0.355\ \\
2001 Jan 16 &  90.60 &  0.349 & 90.72 & 0.409\\
2001 Jan 17 &  91.56 &  0.410 & 91.67 & 0.459\\
2001 Jan 18 &  92.55 &  0.473  & 92.65 & 0.510\\
2001 Jan 19 &  93.56 &  0.537 & 93.68 & 0.564\\

\enddata
\tablenotetext{a}{from \citet{mat97}}

\end{deluxetable}

\clearpage
\begin{figure}

\epsscale{1} \plotone{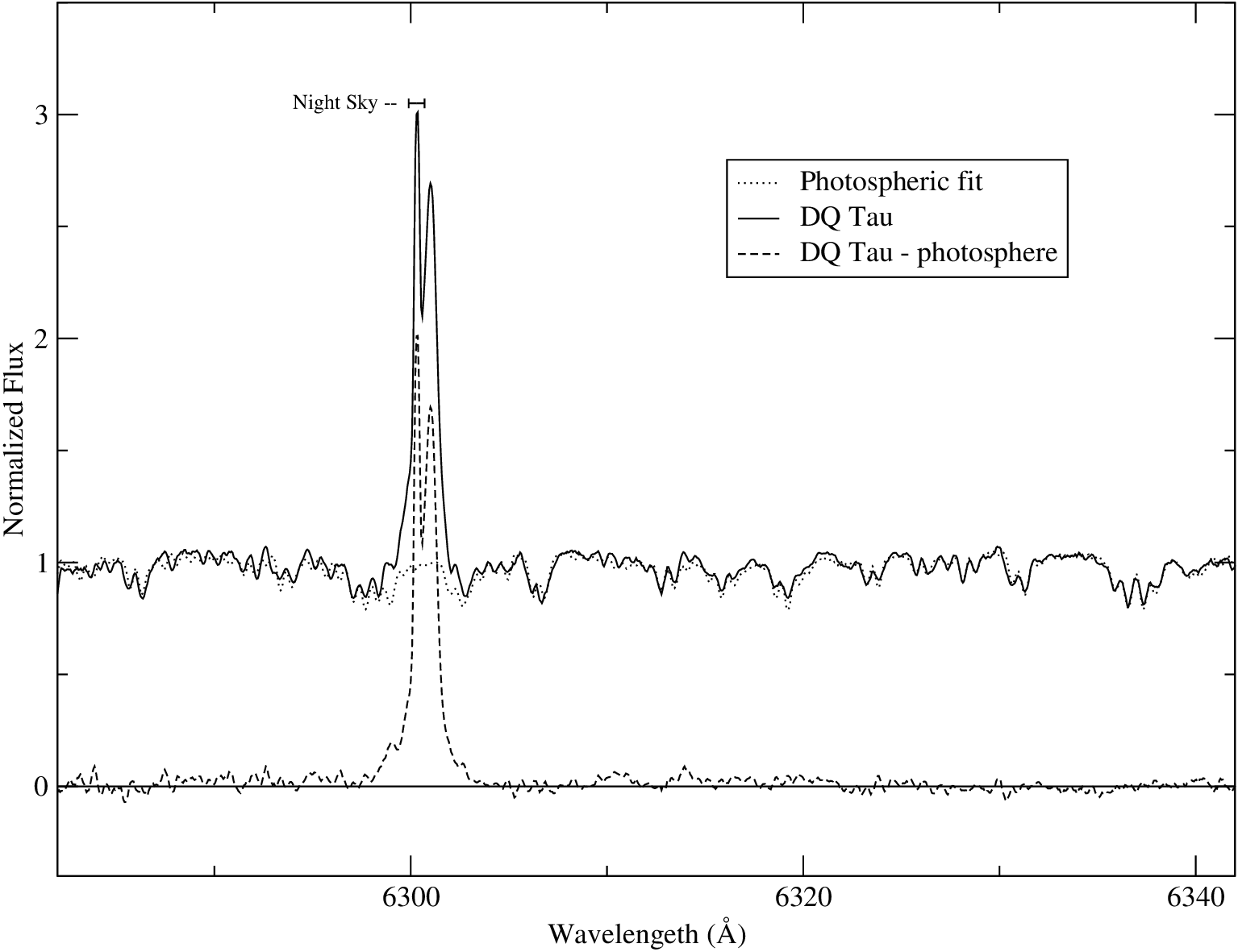}

\caption{Photospheric fit and subtraction for DQ Tau.  The upper curves are the spectra of DQ Tau and the photospheric fit created from the shifted photospheric templates.  The bottom curve is DQ Tau minus the photospheric fit.  The large photospheric features present in the original spectra disappear in the residual spectrum, leaving only the line profile of [O I] $\lambda$6300.  The night sky line is still present in these spectra.  Flux is in units of the stellar continuum.}
\end{figure}
\clearpage

\begin{figure}

\plotone{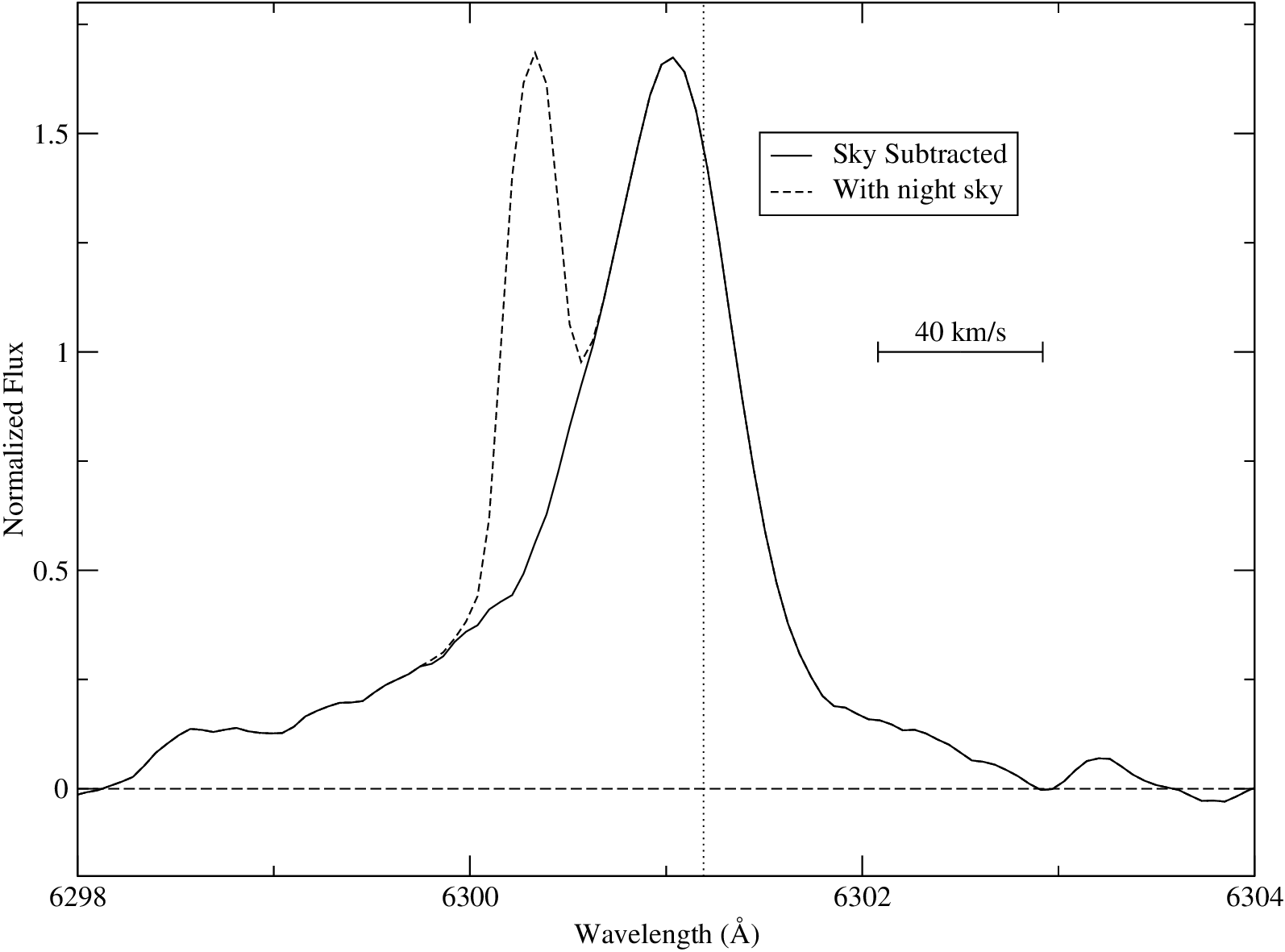}

\caption{Typical before and after sky subtraction for DQ Tau for [O~I] $\lambda$6300.  The original spectrum is present except in the segment near the night sky line, which was replaced with a segment from the sky-subtracted spectrum, as described in section 2.}

\end{figure}

\clearpage

\begin{figure}
\plotone{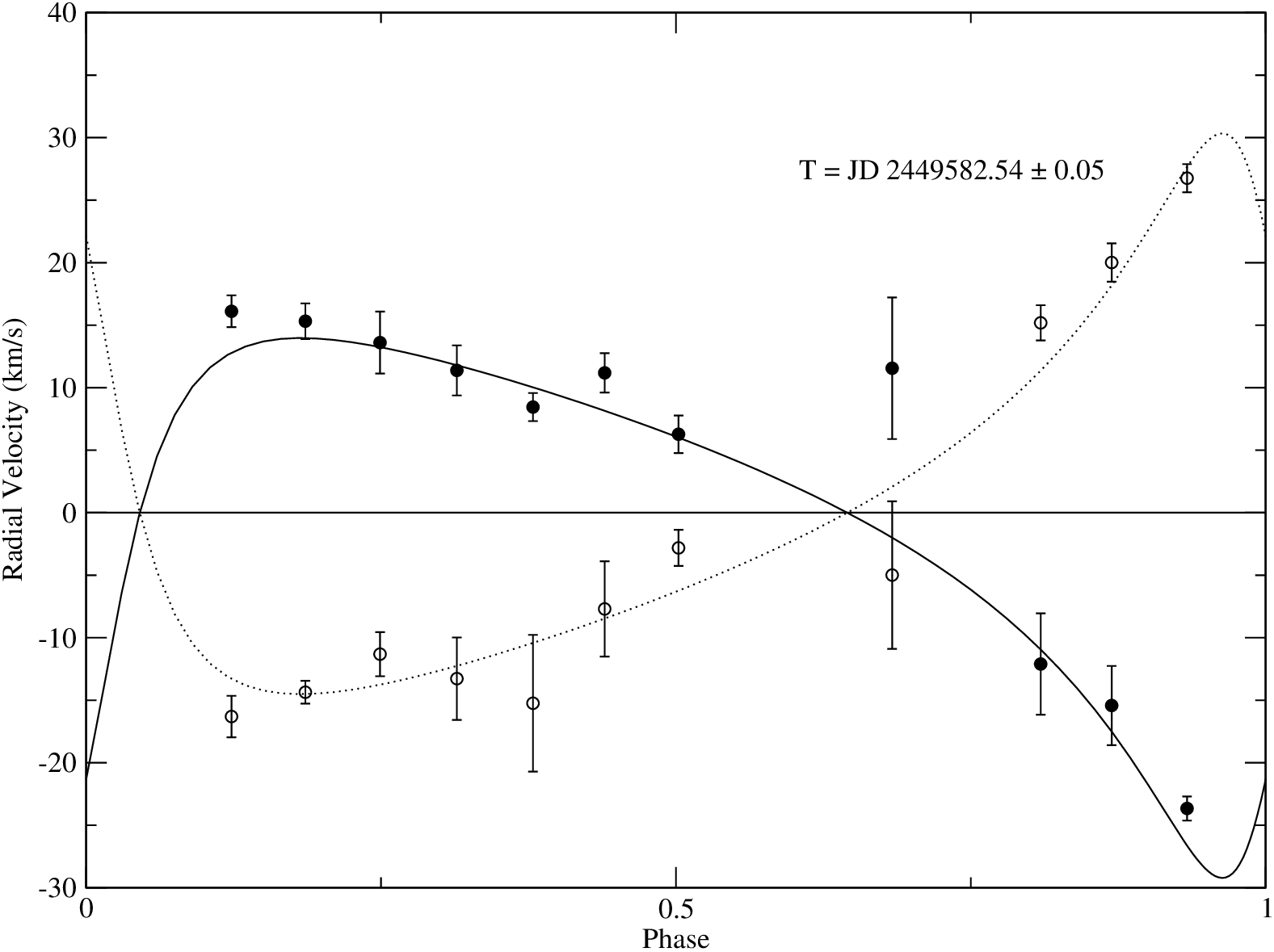}
\caption{Radial velocity versus phase for the primary and secondary in DQ Tau.  Solid circles represent our radial-velocity measurements of the primary, and open circles represent the secondary.  The solid and dashed lines represent orbits for the primary and secondary respectively with our new period value,  P =  15.8016 $\pm^{0.002}_{0.006}$ days, and all other orbital elements taken from \citet{mat97}. }

\end{figure}
 
\clearpage
\begin{figure}

\epsscale{1}
\plotone{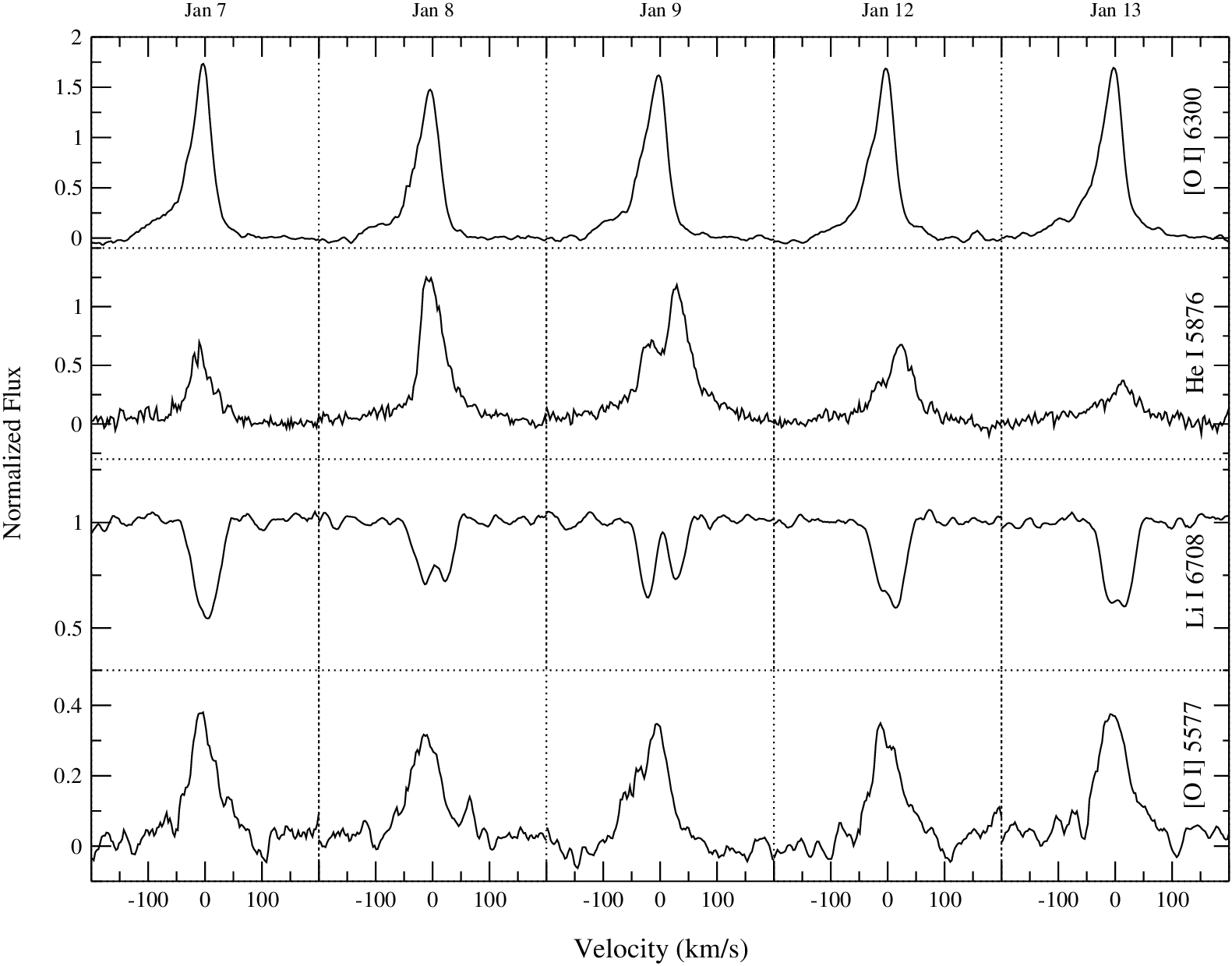}

\caption{Plots for 5 days near periastron for 4 spectral features in DQ Tau.  The flux has been normalized to the continuum.  For the emission line profiles,  the photosphere was subtracted.  The He I line mimics the velocity profile changes of the photospheric lines, while the forbidden lines remain unchanged.}
\end{figure}
\clearpage

\begin{figure}
\epsscale{.8}
\plotone{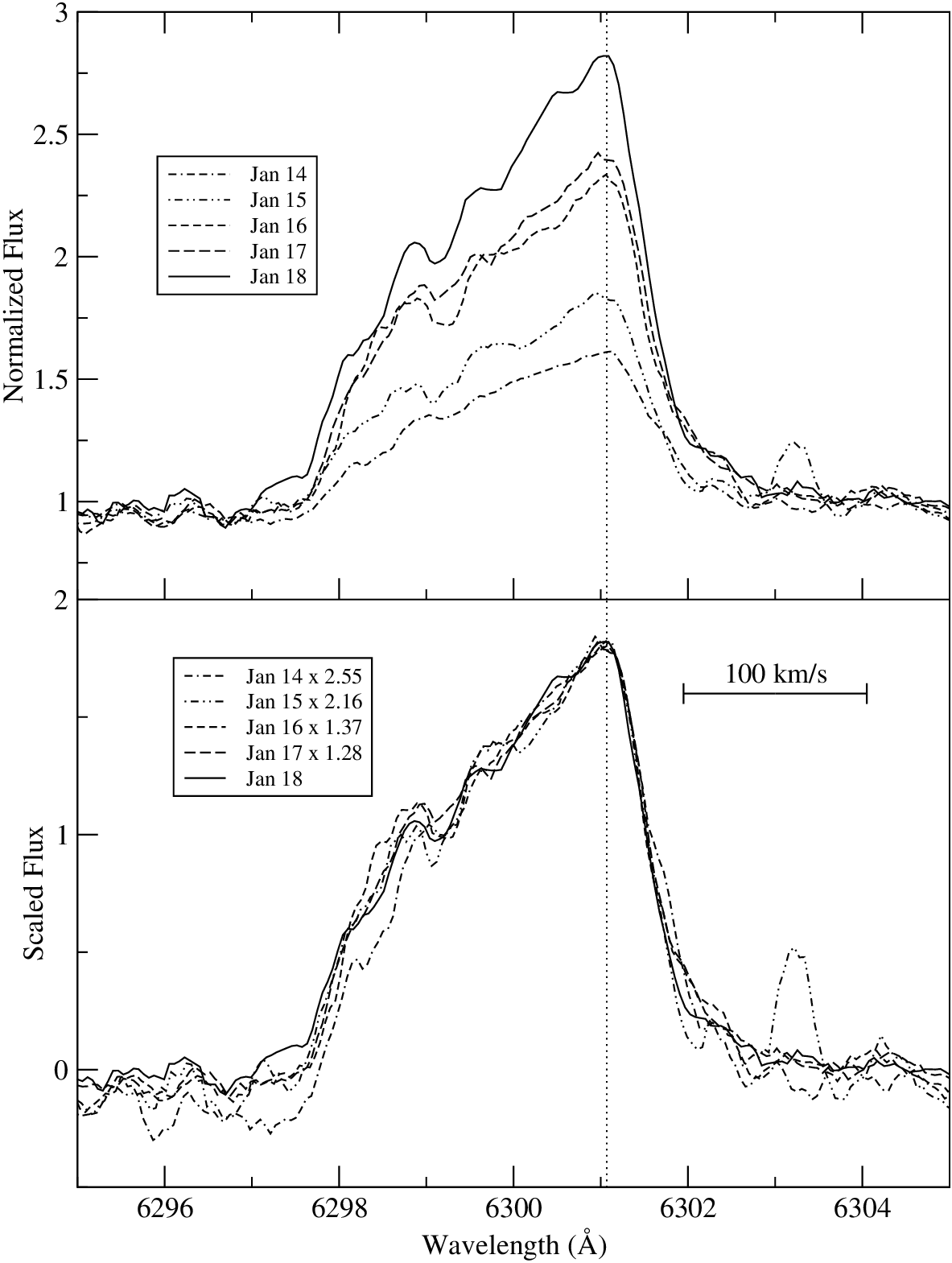}
\caption{Five nights of the [O~I] $\lambda$6300 line in UZ Tau-E.  In the top panel, flux has not been corrected for intensity variations due to veiling, and has been normalized to the continuum.  In the bottom panel, the continuum has been subtracted and each spectrum scaled by the constant shown in the legend, so all have the same peak strength.  The basic line profile shape is unchanged over a time span near periastron.  The vertical doted line indicates velocity of the center of mass of the binary system, from \citet{pra02}.}

\clearpage

\end{figure}

\begin{figure}
\epsscale{1.1}
\plotone{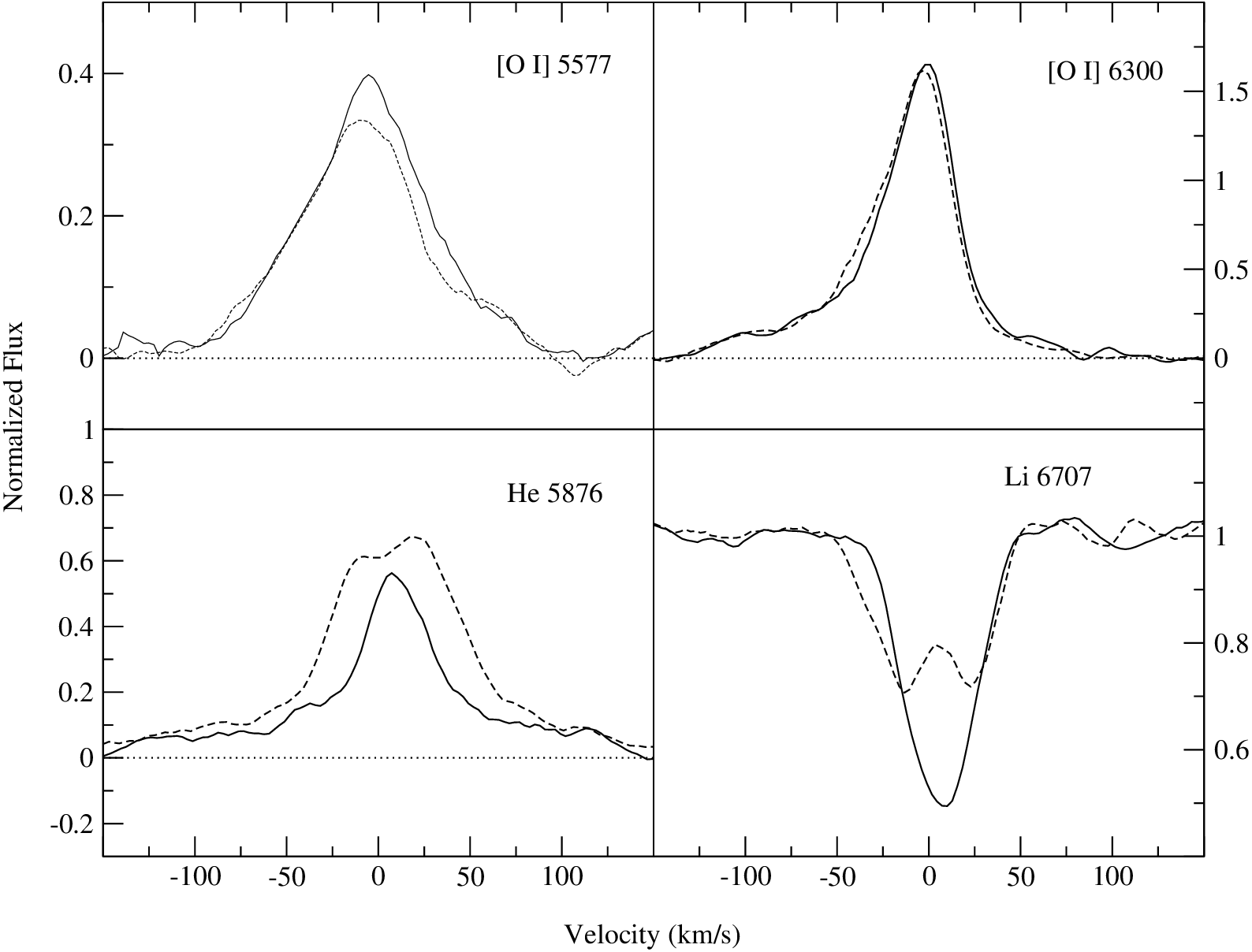}
\caption{Comparison of multi-day averages of emission and absorption lines in DQ Tau.  The [O~I] line profiles show no significant  difference between high-velocity (dashed-lines) and low-velocity (solid lines) averages, unlike He I $\lambda$5876 or Li $\lambda$6707.  The flux of each spectrum has been normalized to its continuum.  In the emission line profiles, the photosphere has been fit and removed.}
\end{figure}

\end{document}